# Programmable recirculating bricks mesh architecture for quantum photonics

Jacek Gosciniak

*Institute of Microelectronics and Optoelectronics, Warsaw University of Technology,*

*Koszykowa 75 st., Warsaw 00-662, Poland*

Email: jacek.gosciniak@pw.edu.pl

**Abstract**

General-purpose programmable photonic processors offer a flexible foundation for integrating various functionalities within a single chip. A two-dimensional hexagonal waveguide mesh of Mach–Zehnder interferometers has been shown to have great potential in the field of microwave photonics. Additionally, they are a promising platform for the creation of unitary linear transformations, which are key elements in photonic neural networks.

In this article, we expand the portfolio of available applications for recirculating "bricks" mesh architecture to quantum technologies. We will show that a single programmable optical system is capable of performing various functions depending on the requirements. In particular, we will focus in this work on boson sampling, a task that best demonstrates quantum advantage, as well as on tasks that enable the determination of photon indistinguishability, which plays a key role in photonic quantum technologies. We will also show that, in addition to spatial modes, the same optical system can be equally well suited for work on temporal modes through the implementation of an appropriate number of loops.

**Introduction**

The rapid progress of quantum computing in the past few years has been primarily driven by the prospect of computational advantage, that is, their ability to solve problems faster than classical computers [1, 2]. Among the milestones, boson sampling emerges as a paradigmatic example demonstrating that a simple, non-universal photonic processor could perform a task believed to be intractable for classical computers [3-7]. Crucial to the development of this promising technology is the generation and manipulation of multiple perfectly indistinguishable photons that enable the implementation of effective photonic gates [8-10]. Integrated photonics is a promising hardware candidate for those tasks and universal fault-tolerant quantum computing forming the backbone of scalable photonic quantum information processing technologies [11, 12]. At the heart of this technology lies a network composed of multiple beam splitters and phase shifters forming multiport interferometers [11, 13]. These multiport interferometers provide precise control over quantum states of light, and their implementation as photonic integrated circuits (PICs) enhances resilience to environmental noise, which is critical for high-fidelity, coherent quantum state manipulation. Such circuits should be characterized by universal reconfigurability enabling a device to be programmed to realize any unitary transformation between input and output modes. The core of the programmable PIC's processor is a photonic waveguide mesh, a two-dimensional lattice that provides regular and periodic geometry, formed by replicating unit cells [14-18]. This is usually achieved using the feed-forward architecture such as triangular mesh architecture proposed by Reck [19] or rectangular mesh architecture proposed by Clements [20], however, the flow of light in those networks is limited only to one direction [21-23]. To open up the entire spectrum of signal processing, there is a need to move to networks that offer far broader possibilities. A recirculating mesh provides significant advantages, primarily in terms of scalability, high reconfigurability and reduced optical losses compared to traditional feed-forward mesh architectures [15, 17, 24]. Instead of building a massive, one-way interferometer, a recirculating mesh allows photons to pass through a smaller, programmable, and tunable component multiple times to simulate a larger, complex unitary transformation.

**Recirculating "bricks" mesh architecture**

To perform any unitary operations between the input and output modes, the interferometric circuits are required. They are arranged in different waveguide mesh architectures able to realize reconfigurable linear transformation upon programming.

Waveguide meshes may generally be classified into two main categories based on signal processing capabilities: feed-forward meshes [**19-23**], where the light flows only in one direction, forward, and recirculating meshes [**14-18, 24**], where light may flow in any available direction, and may be additionally routed in loops and even back to the input ports. In the case of recirculating meshes, they typically adopt a triangular, square, or hexagonal geometry [**18**], where each side of the cell is formed by a tunable basic unit (TBU) composed of one MZI [**18**]. Each such a cell can be programmed and serve as the core of reconfigurable systems, enabling a light flow in both the forward and backward directions. Many such structures comprise the Field Programmable Photonics Gate Array (FPPGA) [**14, 15**] that is analogous to a Field Programmable Gate Array (FPGA) in electronics, offering a unique opportunity to create a flexible platform with versatile functionalities. Such device serves as a core of the photonic chip able to perform multiple operations in different fields ranging from quantum technologies to microwave photonics (neuromorphic computing) [**13, 14**].

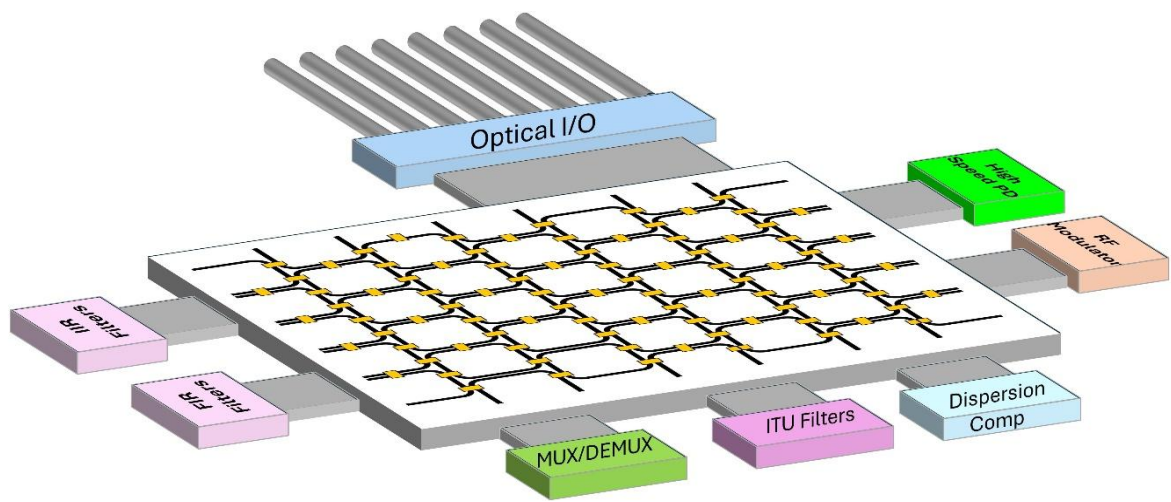

**Figure 1**. The photonic chip with a programmable mesh core connected to control electronics, optical fibers, filters, high-speed modulators and detectors and many other components

Recently, a new type of recirculating mesh architecture was proposed, the so-called "bricks" mesh architecture, which is a slight modification of the standard square mesh architecture [**24**]. This minor modification, however, takes a network built on this type of mesh to a whole new level. The unit cell for this type of mesh consists of from 2 to 4 MZIs, depending on arrangement, compared to 6 for a state-of-the-art hexagonal mesh. In consequence, the optical path length is reduced, which decreases the optical propagation losses and, additionally, enables the synthesis of larger FSR filters. Simultaneously, it yields a more efficient 3-point interconnection scheme compared to a regular square mesh architecture where each unit cell is connected through 4 points.

The configuration with two MZIs is very similar to feed-forward mesh architecture, but unlike it, it allows the signal to be redirected back to its initial port or in any other direction. The information flow is not limited to only one direction, e.g., left-right, up-down, but opens up new possibilities in which all ports can be both input and output ports, as is the case with recirculating meshes. This allows for the implementation of control loops and filters, an infinite impulse response (IIR) filters based on ring resonators (RRs), and a finite impulse response (FIR) based on the asymmetric MZIs, which significantly expands the portfolio of feedforward meshes.

The reconfiguration performance of the mesh, defined as the number of filters with different frequency separation values for the RR-based filter, shows the best performance for this "bricks" mesh, where it was calculated at 11 for only 25 MZIs. In comparison, for a hexagonal mesh and triangular mesh for the same number of MZIs, it was calculated at 9 and 6, respectively, while for a regular square mesh, it was calculated at 6. The reconfiguration performance for the MZI filter with a shifted rectangular/square mesh exceeds 12, which is similar to the performance for the hexagonal mesh evaluated at 12. For a triangular mesh, it was calculated at 8, while for a regular square mesh, at 6. As observed from above, the shifted rectangular/square mesh doubles the performance compared to a regular square mesh architecture.

Another advantage of a proposed recirculating "bricks" mesh architecture is its compatibility with a recently proposed monitoring system that can monitor a signal flow in each part of the circuit under operation [**25**]. Additionally, it can perform a control strategy to automatically adjust the optical power and phase of photonic components at each point of the photonic system with extreme accuracy and minimal insertion loss. The technique is based on a feedback control loop that simultaneously adjusts the matrix coefficient of the device transfer function and compensates for process tolerances and thermal drift in real time. This control strategy relies on the Wheatstone bridge arrangement with a calibration-free feedback loop that does not require prior knowledge of the device transfer function [**25**].

The system can be built based on the transparent conductive oxides (TCOs) that are characterized by the epsilon-near-zero (ENZ) point whose properties can be actively modified by the external voltage [**26-29**].

**Unitary operations**

To implement a desired unitary operation $U$ on-chip, two elementary linear optical components are required - the beam splitter (BS), a constant 2-mode transformation and the phase-shifter (PS), a parameterized 1-mode transformation [**30-33**]. The beam splitter allows mixing between two input electromagnetic modes while the phase shifter adds a phase shift $\phi$ between two modes. The beam splitter and a phase-shifter can build a canonical Mach-Zehnder Interferometer (MZI). In this paper we define MZI with either two phase-shifters placed in both arms of the MZI or with an extra phase-shifter on the input. Each such component creates a two-mode unit cell that can perform a gate operation.

The mode operator $a_i^\dagger$ generated by a unitary evolution $U$ on modes $a_j^\dagger$ is expressed as

$$a_i^\dagger = \sum_{j=1}^{m} U_{ij} a_j^\dagger \qquad (1)$$

The cascaded combination of these two elements can realize any unitary operation in arbitrary dimension. Such decomposition represents the basis for the realization of universal linear optics circuits [**31**].

The unit cell implemented as an MZI with two internal and no external phase shifters is called a symmetric MZI (sMZI). This type of sMZI is attractive because it is more compact without the need for an external phase-shifter, which can account for a significant fraction of the length of the circuit. A shorter structure not only occupies less area on a chip but also suffers from less propagation losses what is especially important in photonic quantum computing. Alternatively, the unit cell implemented as an MZI with an internal phase-shifter in one of the arms and an external phase-shifter on one output is called an asymmetric MZI (aMZI). The internal phase-shifter controls the splitting ratio between the two outputs, while the external controls the relative phase between the outputs. In this way, a universal 2 × 2 gate can be implemented [**21, 34**].

The 50:50 beam splitter can be represented by a matrix

$$U_{BS} = \frac{1}{\sqrt{2}}\begin{bmatrix} 1 & i \\ i & 1 \end{bmatrix} \tag{2}$$

while the phase-shifter by a matrix

$$U_{PS} = \begin{bmatrix} e^{i\phi} & 0 \\ 0 & 1 \end{bmatrix} \tag{3}$$

Thus, the transfer function of the full sMZI (BS-MZI-BS) is expressed as

$$\begin{aligned} U_{sMZI} &= U_{BS}\begin{bmatrix} e^{i\phi_1} & 0 \\ 0 & e^{i\phi_2} \end{bmatrix}U_{BS} = \frac{1}{2}\begin{bmatrix} e^{i\phi_1} - e^{i\phi_2} & i(e^{i\phi_1} + e^{i\phi_2}) \\ i(e^{i\phi_1} + e^{i\phi_2}) & -e^{i\phi_1} + e^{i\phi_2} \end{bmatrix} \\ &= ie^{i\left(\frac{\phi_1+\phi_2}{2}\right)}\begin{bmatrix} \sin\left(\frac{\phi_1-\phi_2}{2}\right) & \cos\left(\frac{\phi_1-\phi_2}{2}\right) \\ \cos\left(\frac{\phi_1-\phi_2}{2}\right) & -\sin\left(\frac{\phi_1-\phi_2}{2}\right) \end{bmatrix} \end{aligned} \tag{4}$$

In comparison, the transfer function of the full aMZI (BS-MZI-BS-MZI) is expressed as

$$U_{aMZI} = \frac{1}{2}U_{BS}\begin{bmatrix} e^{i\phi_2} & 0 \\ 0 & 1 \end{bmatrix}U_{BS}\begin{bmatrix} e^{i\phi_1} & 0 \\ 0 & 1 \end{bmatrix} = ie^{i\frac{\phi_2}{2}}\begin{bmatrix} e^{i\phi_1}\sin\left(\frac{\phi_2}{2}\right) & \cos\left(\frac{\phi_2}{2}\right) \\ e^{i\phi_1}\cos\left(\frac{\phi_2}{2}\right) & -\sin\left(\frac{\phi_2}{2}\right) \end{bmatrix} \tag{5}$$

Reconfigurable universal multiport interferometer can implement any linear transformation between the input and output modes and can work on several optical channels. The Reck scheme [**19**] composed of a triangular mesh can perform such a task using *m(m-1)/2* gates distributed over *2m-3* layers (optical depth of *O(2m-3)*) and *m* parallel modes. In comparison, the rectangular architecture in Clements scheme [**20**] requires the same number of gates, however, realized with only *m* layers, thus reducing an optical depth to *O(m)*. Further reduction of an optical depth to *O(n+log(m))* for layers of MZI with arbitrary connectivity was presented in Ref. [**33**].

Here, the depth of a circuit is defined as the longest path through the circuit, from entrance of exit, enumerated by counting the number of beam splitters traversed by that path.

**Photon loss and photon distinguishability**

The main drawback of photonic quantum computers is their low tolerance to noise. The two main sources of noise can be divided into photon loss and photon distinguishability. Photon loss arises mainly from two sources: coupling losses at the fiber-waveguide interfaces and propagation losses withing the circuit and it is the most physically relevant experimental imperfection in boson sampling. The primary source of photon loss is propagation loss, which is closely dependent on the system architecture and is determined by the optical depth of the chip $D(m)$. In consequence, it depends on the number of components the photons have to go through and on the loss per Mach-Zehnder interferometer (MZI) [**35**]. Thus, minimizing the optical depth of the chip significantly reduces photon loss. Additionally, the problem with photon losses arises as the number $n$ of photons increases what again can play a significant role in the boson sampling experiments.

Another significant source of noise in photonic quantum systems is photon distinguishability which stems from imperfections of the source and from the interaction between the source and the environment. In consequence, it induces changes in internal photon degrees of freedom such as polarization, frequency, and arrival time. As pointed out by Aaronson and Arkhipov in their original paper [**30**], if all photons are completely distinguishable boson sampling becomes classically simulable.

**Boson sampling**

Limited rather by our own perception of the world in which we live, than by physical laws, our thought processes naturally evolve in precisely this direction. And our world is under the universal influence of all-encompassing gravity, which significantly directs our thought. This is precisely the reason why, when we think of boson sampling, the first thing that comes to our minds is a pinball machine or a Galton board, both of which are subject to the law of gravity. While a Galton board or pinball game relies on gravity and classical probabilities to determine a ball's path, boson sampling uses interference of identical photons in a linear optical network, circuit, to produce output distributions that are notoriously hard for classical computers to simulate. This way of perceiving the world has such a profound influence on us that even when we think of a maze, the exit is almost always located on the opposite side of the entrance. Even the optical network is tailored to this way of thinking; it is based on a feed-forward mesh architecture, in which the "input" of photons into the optical network is located on one side, and their "output" on the opposite side, which is monitored by detectors that count how many photons arrive at each output port. However, once we change our way of thinking, we can imagine a maze with exits located on all its sides, or even a box full of particles whose chaotic motion occurs in a two-dimensional plane, and whose detection, sampling, takes place in four directions defined by the side walls of the box. For this architecture, however, a feed-forward mesh architecture is no longer sufficient. We must move to a mesh architecture that allows light to propagate in any possible direction, without limiting it to any specific direction. And such a network already exists; it is called the recirculating mesh architecture. Furthermore, it is programmable, thus meeting all the requirements set for such networks. Among these types of networks, one in particular deserves special attention due to its similarity to networks operating on the basis of feed-forward mesh architecture, which is already well known. This network is based on the so-called bricks mesh architecture, also known as shifted rectangular mesh architecture. More information about its advantages can be found in Ref. [**24**].

After this brief introduction, let's now return to discussing what boson sampling is. Boson sampling is a specialized computational task that serves as a benchmark for quantum supremacy. It was designed to be solved easily by a quantum machine but to be practically impossible for even the world's most powerful classical supercomputers. Boson sampling involves (is the task of) sampling from the output distribution of an $m$ mode interferometer implementing a Haar random unitary described by an $m \times m$ unitary matrix $U$ with $n$ input photons. The unitary matrix $U$ represents the linear optical network through which the photons propagate. There is ample evidence that performing boson sampling on $n > 50$ photons with $n \ll m$ is sufficient to reach quantum utility [**4**]. The key idea behind boson sampling is the connection between the output statistics of non-interacting indistinguishable photons and the permanent, a quantity known to be #P-hard to compute by classical computers [**3**].

Let assume $m$ is a set of modes associated with creation operators $a_i^\dagger$, where $i = 1, \ldots, m$, and $S = \{s_1, s_2, \ldots, s_m\}$ is the occupation-number tuple that represents a possible input configuration of the photons. Here, $s_i \in \{0, \ldots, n\}$ are the non-negative integers that represent the number of photons in each mode $i$ where $\sum_{i=1}^{m} s_i = n$. For $s_i \leq 1$, each state is a non-collision state meaning that there is no individual mode containing more than one photon. Considering now an $m$-mode linear optical transformation $U \in SU(m)$, the evolution of the creation operator is expressed by *eq. 1.* while a Fock state of $n$ photons in modes $m$ can be written as

$$|S\rangle = |s_1 s_2 \dots s_m\rangle = \prod_{i=1}^{m} \frac{\left(a_i^\dagger\right)^{s_i}}{s_i!}|0\rangle. \tag{6}$$

In the case of boson sampling, the task involves sampling from the distribution over output configuration $V = \{v_1, v_2, \dots, v_m\}$, where $v_i \in \{0, \dots, n\}$ and the number of photons is conserved, i.e., $\sum_{i=1}^{m} v_i = n$. Thus, the transmission probability between an input state $|S\rangle = |s_1 s_2 \dots s_m\rangle$ and an output state $|V\rangle = |v_1 v_2 \dots v_m\rangle$ can be written as:

$$P_{s,v} = \frac{|\text{Per} U_{s,v}|^2}{s_1! s_2! \dots s_m! v_1! v_2! \dots v_m!} \tag{7}$$

where the probability of observing a specific output configuration $V$ given an input configuration $S$ is related to the permanent of a matrix $U_{s,v}$. Here, $U_{s,v}$ is an $n \times n$ submatrix of $U$ ($m \times m$) constructed by selecting rows and columns of the matrix $U$ describing the linear transformation, and permanent of a $n \times n$ matrix $B$ is expressed as

$$\text{Per}(B) = \sum_{\sigma \in S_n} \prod_{i=1}^{n} b_{i,\sigma(i)} \tag{8}$$

where $S_n$ is the symmetric group of dimension $m$, and $\sigma$ represent a generic permutation [**4, 35**].

In general, the permanent is hard to compute by classical algorithms as a computation requires $O(n2^n)$ steps for an $n \times n$ matrix with complex entries and grows exponentially with the number of photons $n$. It requires the brute-force approach that involves calculation of all probabilities associated with all input-output possibilities. In consequence, it belongs to the #P-hard class in complexity theory [**36**].

Thus, the larger the rank of the unitary matrix $U$, the larger the permanent and the more complex the system. However, as the complexity of the system increases, the photon loss also arises, which is primarily due to an increase in the optical depth of the system; consequently, minimizing photon loss while simultaneously increasing the complexity of the system is one of the objectives in boson sampling.

Compared to any other mesh architecture, in which optical signal flow is essentially limited to a single direction—and which therefore falls under the category of so-called feed-forward mesh architectures, in the recirculating meshes the light can be routed in any possible direction, in the loops and even back to the input port. This opens up a whole new field in the development of programmable photonic circuits with unprecedented performance, in which a single programmable circuit can perform multiple tasks simultaneously. Additionally, signal processing is not limited solely to spatial signals, as is the case of feed-forward meshes, but also extends to temporal signals, which brings some additional degree of freedom to the system. This enables the implementation of circuits that are computationally universal, capable of performing any linear optical unitary operation while minimizing photon loss and gate infidelity. A simple approach to minimizing photon loss is to reduce the size and optical depth of the circuit, while ensuring the implementation of all necessary signal processing operations. Since signal processing errors and fidelity errors, which arise due to noise and cross-talk in phase settings, are mainly caused by inaccuracies in the implementation of active circuit devices such as phase-shifters, a simple approach to minimize such errors involves minimizing the total number of phase-shifters. The proposed configuration, based on a "bricks" mesh architecture, allows these two goals to be achieved by minimizing both the optical depth of the chip and the number of phase-shifters while simultaneously performing all necessary operations.

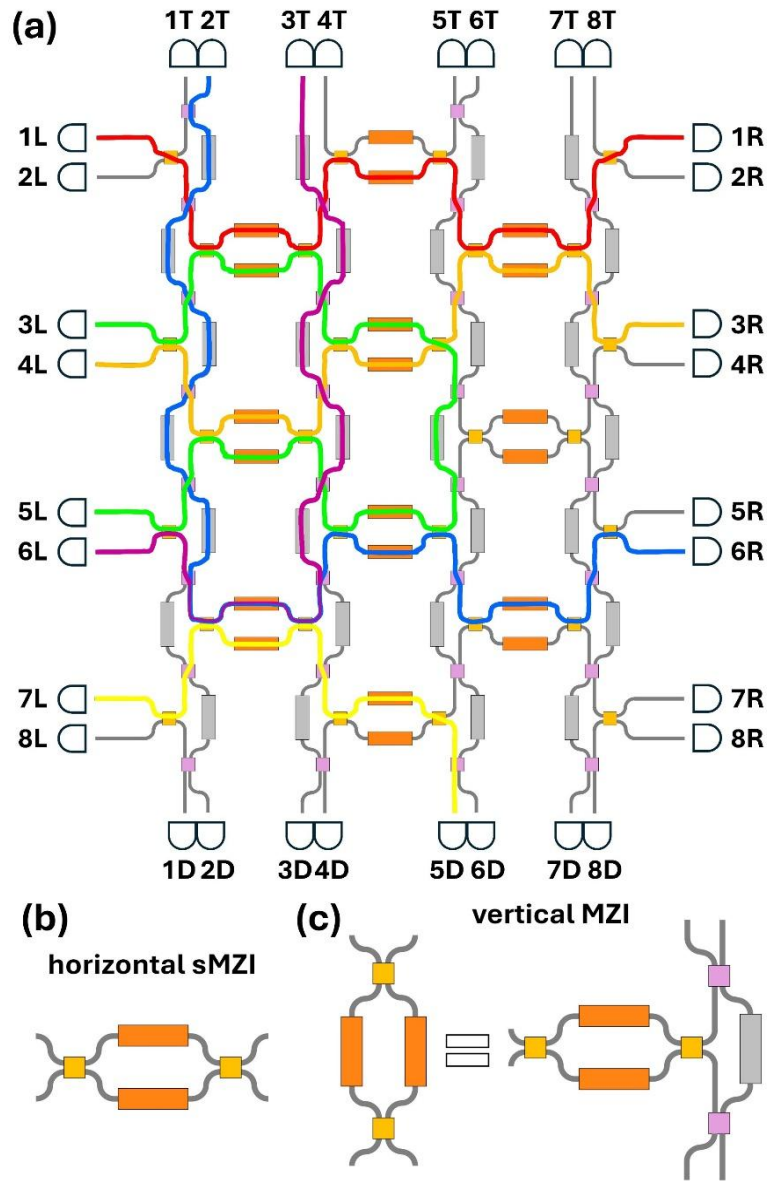

**Figure 2**. A diagram of a "bricks" mesh architecture with several possible paths of light propagation. Horizontal connections are implemented using symmetric MZIs (sMZI) (b), while vertical connections are implemented using modified MZIs (c).

As light can be routed in any arbitrary direction, in loops, and even return back to the input port, the number of modes $m$ available for detection increases four times while the number of gates is kept similar to the feed-forward architectures. An example of "bricks" mesh with $m=32$ modes is shown in **Fig. 2**. It consists of 10 symmetric MZIs arranged in a horizontal direction and 28 modified "symmetric" MZIs arranged in a vertical direction giving an overall number of 38 MZIs. By comparison, simple calculations show that the number of gates, MZIs, for $m=32$ modes implemented in a feed-forward architecture is approximately 496, which is more than 13 times higher than that for a recirculating "bricks" mesh architecture with the same number of modes. A smaller number of active components on the chip capable of performing similar tasks translates directly into a massive reduction in propagation losses, which are the main factor affecting the performance of photon-based quantum computing. This is also of great importance in boson sampling experiments, in which increasing the size and depth of the circuit ultimately leads to a system that is equivalent to sampling random noise.

The increased number of modes $m$ available for detection in the case of "bricks" mesh architecture comes with an increased number of detectors. This translates into a significant increase in the complexity of boson sampling, which depends on the number of detectors triggered rather than on the number of input photons $n$ [**35**].

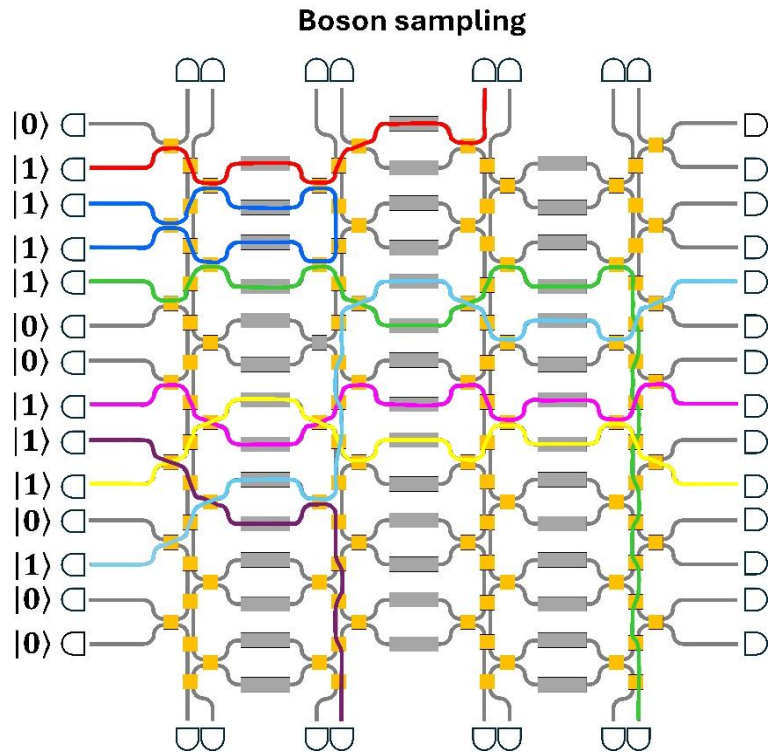

**Figure 3**. A simplified scheme of a recirculating "brick" mesh architecture with several possible light propagation paths. Due to the network architecture, an optical signal can be detected in any direction.

The footprint and the number of components required for more complex circuit operating with higher number of modes scales exponentially. The circuit shown in **Fig. 3**, operating at $m$=44 modes, consists of 21 symmetric MZIs arranged in a horizontal direction and 42 modified MZIs arranged in a vertical direction providing an overall number of 63 MZIs. Consequently, 63 MZIs provide support for all 44 modes, signifying that in the context of a recirculating "bricks" mesh architecture, different modes are consistently supported by at least a subset of the same MZIs. In comparison, a feed-forward architecture operating with the same number of modes will require 946 gates, MZIs, which is 15 times higher than for a recirculating "bricks" mesh architecture operating with the same number of modes. This represents a significant improvement in terms of resource requirements, both in terms of circuit footprint and energy costs as well as the overall losses. The circumstances appear even more favorable when we take into account that, in "bricks" mesh architecture, a signal can be transmitted through each MZI component multiple times.

**Cyclic interferometer for n-photon indistinguishability**

A second key challenge, asides of losses, in the development of photonic quantum technologies is the generation and manipulation of multiple perfectly indistinguishable photons. The reference benchmark for indistinguishability is the Hong-Ou-Mandel (HOM) dip where the level of indistinguishability between two photons can be determined from a binary interference pattern. However, this has been shown to be insufficient for characterizing multi-photon interference [9].

To determine the upper and lower bounds on the probability that *n* photons are completely indistinguishable, a metric known as genuine n-photon indistinguishability (GI) has been introduced. The genuine n-photon indistinguishability is defined as the probability that the n photons are identical. One of the methods for computing GI relies on a cyclic integrated interferometer (CI), whose general layout is presented in **Fig. 4**, and a post-selection strategy that homogenizes the output probability for any (partially) distinguishable input state as soon as there is at least one distinguishable photon at input. This creates a multi-photon HOM-like interference pattern for their post-selected output configurations allowing them to extract GI [8].

In **Fig. 4**, a CI with $m = 2n$ modes is presented, in which $n$=3 photons are injected one per each odd input mode, and $n$-photon coincidence detection is performed at the output, detecting one photon per output modes 2, 3 and 6.

For perfectly identical $n$ photons, the probability to detect this output state is given by [8]

$$P_{s,v} = \frac{|\text{Per}S_{s,v}|^2}{s!\,s_2!\cdots s!\,v_1!\,v_2!\cdots v_m!} \qquad (9)$$

where $s_i$ and $v_i$ are the number of photons (particles) present in mode $i$ in the input state $|s\rangle$ and output state $|v\rangle$, respectively. Here, $S_{s,v}$ is the scattering matrix with elements $S_{i,j} = U_{s_i,v_i}$, where $U$ is the unitary matrix of the interferometer, and $\text{Per}S$ denotes the permanent of the matrix $S$.

The elements of the $n \times n$ scattering matrix $S$ can be calculated by inspection of the interferometer layout, considering the attenuation and phase delays undergone by a photon in all the relevant optical paths.

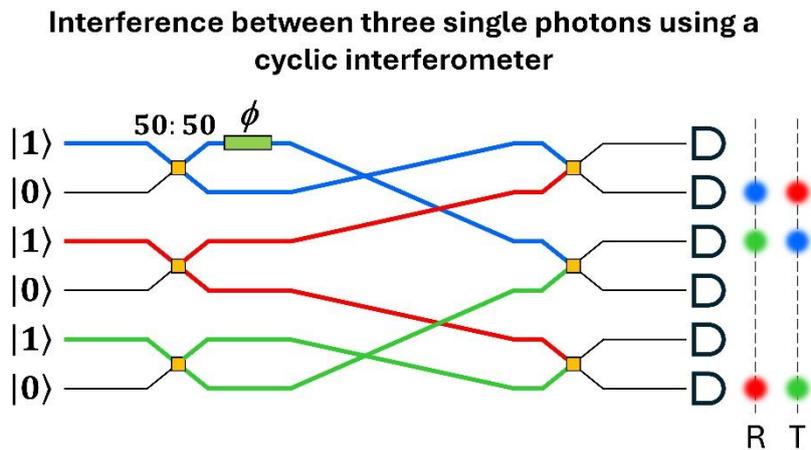

**Figure 4**. Interference between three single photons using a cyclic interferometer (CI) tuned using a single phase-shifter implementing the phase $\phi$.

A photon injected in the 3$^{rd}$ mode can reach the 2$^{nd}$ mode undergoing two 50% transmission on the first and second beam splitters, each giving a phase delay equal to $i = e^{i\pi/2}\,(= cos(\pi/2) + i \cdot sin(\pi/2))$. Thus,

$$S_{3,2} = \frac{1}{2}i^2 = -\frac{1}{2} \qquad (10)$$

Alternatively, a photon from the same input 3rd, may exit from the 6th mode undergoing two 50% reflections, each giving a null phase delay:

$$S_{3,6} = \frac{1}{2}i^2 = \frac{1}{2} \qquad (11)$$

All other elements of the scattering matrix associated to this input are vanishing, because there are no other modes connected to it.

Similar calculation we make for two other odd input modes. For the 1st mode we receive:

$$S_{1,2} = \frac{1}{2}i^2 = \frac{1}{2}, \quad S_{1,3} = \frac{1}{2}i^2 = -\frac{1}{2}e^{i\phi} \qquad (12)$$

while for the 5th mode:

$$S_{5,3} = \frac{1}{2}, \quad S_{5,6} = -\frac{1}{2} \qquad (13)$$

In this way, the entire scattering matrix can be constructed.

Here, for input state $|s\rangle = |101010\rangle$ of $n = 3$ identical photons, corresponding to the combination $s = \{1,3,5\}$, the probability to detect the output state $|v\rangle = |011001\rangle$, corresponding to the combination $v = \{2,3,6\}$ is given by:

$$P = \frac{1}{2^{2n-1}}[1 + (-1)^{n+p+q}cos\phi] \qquad (14)$$

where $p$ is the number of occupied even modes in the input state and $q$ is the number of occupied even modes in the output state.

This specific combination has zero occupied even modes in the input state $|s\rangle$ and two occupied even modes (the modes 2 and 6) in the output state $|v\rangle$. Thus, for $n = 3$ one has $(-1)^{3+0+2} = -1$ and the term $cos\phi$ in the above equation has to be taken with the minus sign

$$P = \frac{1}{2^5}[1 - cos\phi] \qquad (15)$$

where $\phi$ corresponds to either a single internal phase as in **Fig. 4** or a phase delay $\phi = \sum_{m=1}^{n/2} \varphi_{2m-1} - \varphi_{2m}$ placed on odd and even modes that can be implemented through layer 4th as presented in **Fig. 5**.

As shown above, when $n$ indistinguishable photons are injected into the interferometer, the output distribution exhibit quantum interference that depends on either a single internal phase (**Fig. 4**) or a phase from one of the phase-shifters in layer 4 (**Fig. 5**), and the interference visibility corresponds to the genuine $n$-photon indistinguishability [**8**]. Conversely, no quantum interference is observed for input Fock states with less than $n$ photons.

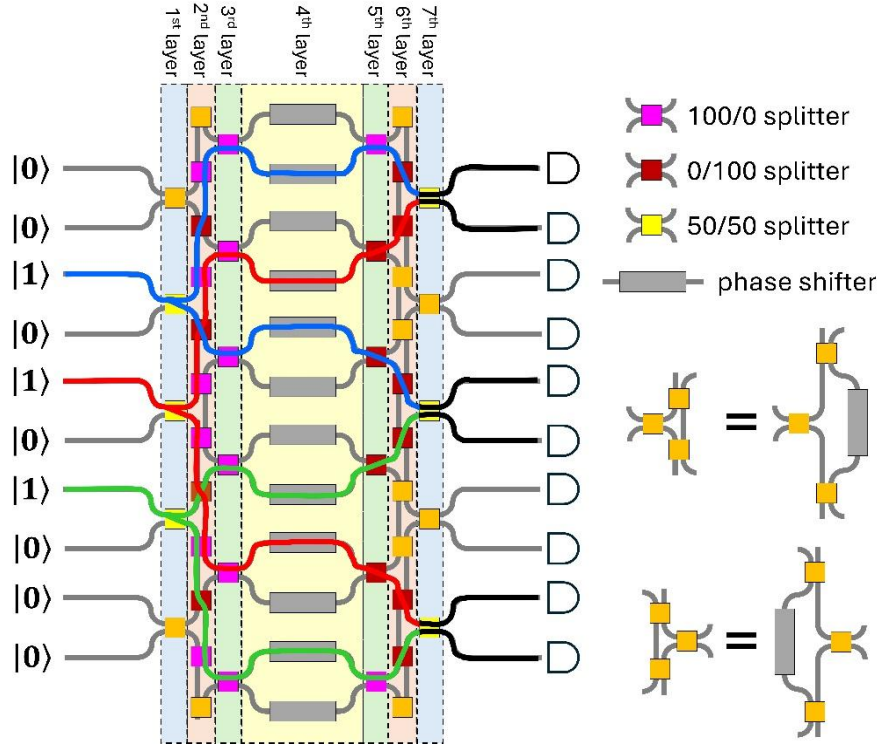

**Figure 5**. Cyclic interferometer (IC) implemented in the recirculating "bricks" mesh architecture.

A scheme of the proposed programmable photonic circuit based on a "bricks" mesh architecture is shown in **Fig. 5**. The odd input ports of the interferometers are fed with single photons. The first layer of balanced beam splitters split the input state composed of *n* photons into two alternative paths while the final layer of balanced beam splitters, 7$^{th}$ layer, connects each two adjacent modes through the phenomenon of interference. Connections between the first and last layers of balanced beam splitters are implemented using intermediate programmable beam splitters. Connections yield cyclic symmetry, as the *nth* photon (here in green color) is interfered with the first one (here in red color). By performing collective coincidence measurements at the output, it is possible to infer an intrinsic property of the whole set of photons. The output distribution for all output ports corresponds to constructive and destructive *n*-photon interference where the visibility of the interference fringes defines the genuine *n*-photon indistinguishability. To achieve the full interference fringe visibility for *n*-photon interferences the internal phase of the interferometer ($\phi$ in **Fig. 4** or any phase shifters from layer 4$^{th}$ in **Fig. 5**) can be scanned. This allows to compute the associated unitary matrix for each value of the internal phase $\phi$ and then transpile the circuit to be implemented on the quantum processing unit (QPU) [**37**]. The coincidence measurements performed at the output enable inferring an intrinsic property of the whole set of photons, i.e., their genuine indistinguishability.

Compared to a standard configuration, the proposed programmable photonic circuit can be implemented in a two-dimensional plane, which significantly simplifies the fabrication and integration process. In standard arrangement, as presented in **Fig. 4**, two waveguides are required to pass one over the others without crossing, to reach the last beam splitter connecting in this way modes 1 and $N$ (here mode 1 and 3). This is possible, however, requires three-dimensional integration.

**Temporal modes with loops**

As it was mentioned above, the central challenge in boson sampling task, experiments, is construction of linear optical networks. It was additionally shown that any arbitrary network of the forms presented above relies on the use of spatial modes where the network can be decomposed into a sequence of beamsplitters.

A fundamentally different approach relies on the use of temporal modes where a train of $M$ time bins, each loaded with one or zero photon and separated by an interval $\tau$, are injected into a loop and circulated for $N$ loops [**1, 38, 39**]. Such a loop-based architecture is equivalent to an $M$-mode beam splitter network with a depth of $N$. In this scenario, the unitary transformation can be implemented by means of two different low-loss loops where the photons can be routed. The smaller loop introduces a delay of $\tau$ for incoming photons, enabling interference between adjacent time bins. In contrast, the larger loop introduces a delay that is significantly greater than τ and feeds the circuit with the same photon packets, thereby enabling subsequent steps of interference.

This scheme can be easily implemented using the same recirculating "bricks" mesh architecture as proposed in this paper by a proper programing the corresponding gates (**Fig. 6**).

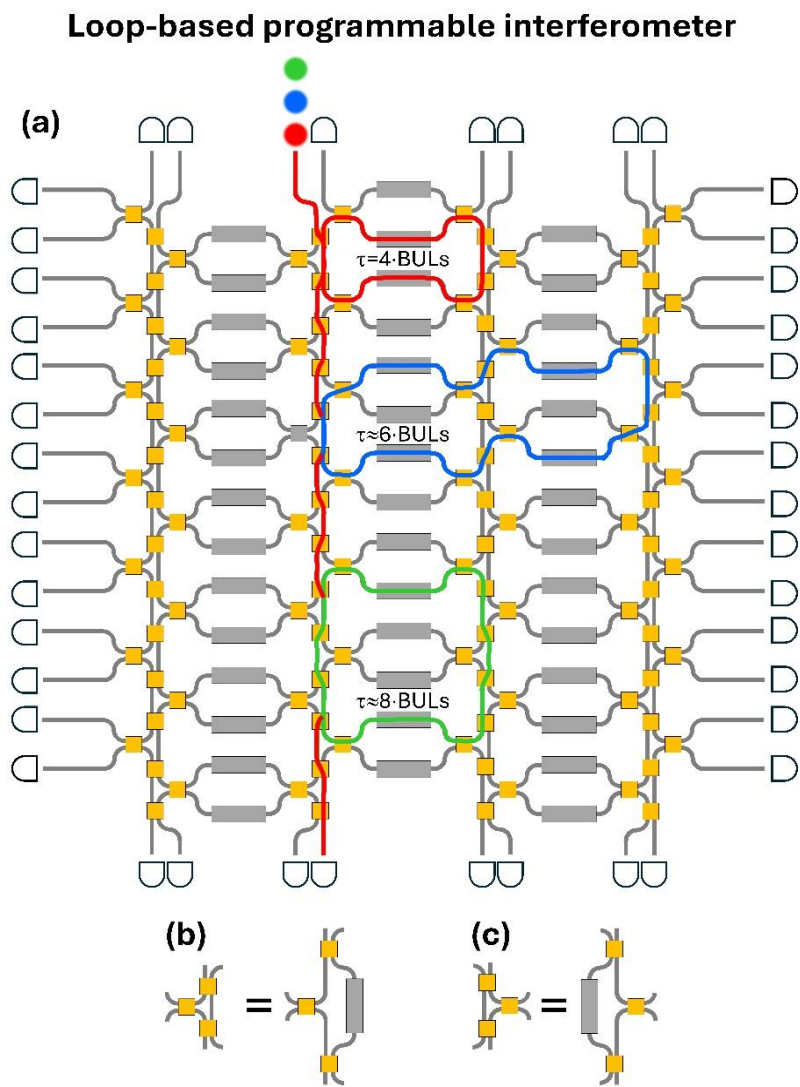

**Figure 6**. A simplified scheme of a recirculating "bricks" mesh architecture to perform some basic operations on temporal modes through the implementation of an appropriate number of loops.

For recirculating mesh networks, the cavity size is essentially determined by the size of the phase shifter and the two beam splitters that together form a gate and constitute the basic unit that defines the size of the unit cell, so called the basic unit length (BUL). Therefore, the minimum cavity size is 4 BULs and is mostly limited by the length of the MZI. Assuming the MZI length at $L_s$=450 µm the corresponding round trip time through a cavity can be calculated at 30 ps, resulting in a spectral period of 34 GHz. In this type of mesh, higher cavities follow in increments of 2 BULs.

Thus, the proposed recirculating "bricks" mesh architecture enables implementation of a variety of functionalities by suitable programming of gates.

**Summary**

Recirculating meshes, and "bricks" meshes in particular, provides significant advantages in quantum technologies, primarily focused on scalability, high reconfigurability, and reduced optical losses compared to traditional, static, large-scale photonic circuits. Instead of building a massive, one-way interferometer, a recirculating mesh allows photons to pass through a smaller, programmable, and tunable component multiple times to simulate a larger, complex unitary transformation.

These systems can be programmed to implement a vast, diverse set of Haar-random unitary transformations, enabling the exploration of different quantum states within the same hardware, which is crucial for testing various quantum technologies.

As recirculating meshes require fewer physical components to construct large-scale interferometers, they allow the number of modes and photons to be scaled up without needing a proportional increase in component footprint. This also results in a reduction in propagation losses, which are typical for feed-forward networks, which is vital for maintaining the quantum nature of the signal. These advantages make "bricks" mesh designs a very promising technology for implementing practical, near-term, large-scale quantum signal processing.

**Acknowledgement**

The author acknowledges the constant support of Warsaw University of Technology, Poland, for the completion of this work. Furthermore, he is very thankful to Prof. D. G. Misiek for his support and very valuable suggestions.